\documentclass{article}
\usepackage{tenor2024}
\usepackage{ifpdf}
\usepackage[english]{babel}
\usepackage{balance}
\usepackage{comment}

\usepackage{amsmath} 
\usepackage{amssymb} 
\usepackage{amsfonts}
\usepackage{bm}      
\usepackage{paralist}

\usepackage[hang,small,bf]{caption} 
\usepackage[subrefformat=parens]{subcaption}

\usepackage{amsthm}  
\theoremstyle{definition}
\newtheorem{example}{Example}[section]

\usepackage{url}

\usepackage{color}

\usepackage{csvsimple}

\usepackage{multicol}

\def\papertitle{Engraving Oriented Joint Estimation of Pitch Spelling and Local and Global Keys}

\def\firstauthor{Augustin Bouquillard}
\def\secondauthor{Florent Jacquemard}

\newif\ifpdf
\ifx\pdfoutput\relax
\else
   \ifcase\pdfoutput
      \pdffalse	
   \else
      \pdftrue
\fi

\ifpdf 
  \usepackage[pdftex,
    pdftitle={\papertitle},
    pdfauthor={\firstauthor, \secondauthor},  
    bookmarksnumbered, 
    pdfstartview=XYZ 
   ]{hyperref}

  \usepackage[pdftex]{graphicx}
  \graphicspath{{./figures/}}
  \DeclareGraphicsExtensions{.pdf,.jpeg,.png}

  \usepackage[figure,table]{hypcap}

\else 
  \usepackage[dvips,
    bookmarksnumbered, 
    pdfstartview=XYZ 
  ]{hyperref}  

  \usepackage[dvips]{epsfig,graphicx}
  \graphicspath{{./figures/}}
  \DeclareGraphicsExtensions{.eps}

  \usepackage[figure,table]{hypcap}
\fi

\hypersetup{
    colorlinks,%
    citecolor=black,%
    filecolor=black,%
    linkcolor=black,%
    urlcolor=black
}

\makeatletter
\newcommand\footnoteref[1]{\protected@xdef\@thefnmark{\ref{#1}}\@footnotemark}
\makeatother

\usepackage[disable]{todonotes} 
\setuptodonotes{noline,size=\tiny}
\newcommand{\florent}[1]{\todo[color=yellow!40]{#1}}

\usepackage{xspace}

\usepackage{musicography}

\def\<#1>{\langle #1 \rangle}

\def\Ap{\mathsf{A}}
\def\Bp{\mathsf{B}}
\def\Cp{\mathsf{C}}
\def\Dp{\mathsf{D}}
\def\Ep{\mathsf{E}}
\def\Fp{\mathsf{F}}
\def\Gp{\mathsf{G}}

\def\Ass{\mathsf{A}\musDoubleSharp}
\def\Bss{\mathsf{B}\musDoubleSharp}
\def\Css{\mathsf{C}\musDoubleSharp}
\def\Dss{\mathsf{D}\musDoubleSharp}
\def\Ess{\mathsf{E}\musDoubleSharp}
\def\Fss{\mathsf{F}\musDoubleSharp}
\def\Gss{\mathsf{G}\musDoubleSharp}

\def\As{\mathsf{A}\musSharp} 
\def\Bs{\mathsf{B}\musSharp}
\def\Cs{\mathsf{C}\musSharp}
\def\Ds{\mathsf{D}\musSharp}
\def\Es{\mathsf{E}\musSharp}
\def\Fs{\mathsf{F}\musSharp}
\def\Gs{\mathsf{G}\musSharp}

\def\An{\mathsf{A}\musNatural} 

\def\Cn{\mathsf{C}\musNatural}
\def\Dn{\mathsf{D}\musNatural}

\def\Fn{\mathsf{F}\musNatural}
\def\Gn{\mathsf{G}\musNatural}

\def\Af{\mathsf{A}\musFlat} 
\def\Bf{\mathsf{B}\musFlat}
\def\Cf{\mathsf{C}\musFlat}
\def\Df{\mathsf{D}\musFlat}
\def\Ef{\mathsf{E}\musFlat}
\def\Ff{\mathsf{F}\musFlat}
\def\Gf{\mathsf{G}\musFlat}

\def\Aff{\mathsf{A}\musDoubleFlat}
\def\Bff{\mathsf{B}\musDoubleFlat}
\def\Cff{\mathsf{C}\musDoubleFlat}
\def\Dff{\mathsf{D}\musDoubleFlat}
\def\Eff{\mathsf{E}\musDoubleFlat}
\def\Fff{\mathsf{F}\musDoubleFlat}
\def\Gff{\mathsf{G}\musDoubleFlat}

\newcommand{\pc}{\mathit{pc}}

\def\ie{\textit{i.e.},\xspace}
\def\eg{\textit{e.g.},\xspace}

\def\etc{\textit{etc}\xspace}

\title{\papertitle}

\twoauthors
  {\firstauthor} {Université Paris Dauphine, PSL \\ %
    {\tt \href{mailto:augustinus.bouquillard@orange.fr}{augustinus.bouquillard@orange.fr}}}
  {\secondauthor} {Inria Paris and Cedric/CNAM \\ %
    {\tt \href{mailto:florent.jacquemard@inria.fr}{florent.jacquemard@inria.fr}}}

\begin{document}
\capstartfalse
\maketitle
\capstarttrue
\begin{abstract}
We revisit the problems of pitch spelling and tonality guessing 
with a new algorithm for their joint estimation
from a MIDI file including information about the measure boundaries.
Our algorithm
does not only identify a global key but also local ones all along the analyzed piece.
It uses Dynamic Programming techniques to 
search for an optimal spelling in term, roughly, 
of the number of accidental symbols that would be displayed in the engraved score.
The evaluation of this number is coupled with an estimation of the global key 
and some local keys, one for each measure.
Each of the three informations is used for the estimation of the other, 
in a multi-steps procedure.

An evaluation conducted on a monophonic and a piano dataset, comprising 216 464 notes in total, shows a 
high degree of accuracy, 
both for pitch spelling 
(99.5\% on average on the Bach corpus and 98.2\% on the whole dataset) 
and global key signature estimation (93.0\% on average, 
95.58\% on the piano dataset).

Designed originally as a backend tool in a music transcription framework, 
this method should also be useful in other tasks related to music notation processing. 
\end{abstract}


\maketitle

\section{Introduction}\label{sec:introduction}
%
In symbolic music representations, pitches are expressed in different ways.
In their simplest form, in the MIDI standard,
they are encoded as integers corresponding to a number of keys on a device.
The representation of pitches is much more involved
in Common Western Music Notation (CWN), 
where the denotation of each note depends on the musical context 
of occurrence:
the tonality (key) of the piece,
the voice-leading structure (ascending or descending melodic movements), 
the harmonic context...

If we reason modulo octaves, every pitch class, 
between 0 and 11 semitons, 
can be denoted in several ways, 
using a note name, in $\Cp$, $\Dp$, $\Ep$, $\Fp$, $\Gp$, $\Ap$, $\Bp$,
and an accidental mark in 
\musDoubleFlat, 
\musFlat, 
\musNatural, 
\musSharp, 
\musDoubleSharp, 
acting as a pitch class modifier.
\emph{Pitch-Spelling} (PS) is the problem of choosing appropriate names
to denote some given MIDI pitch values in CWN.

\begin{figure}
\centering
\begin{subfigure}{0.45\textwidth}
\centering
\includegraphics[height=0.1\textheight,width=0.8\textwidth]{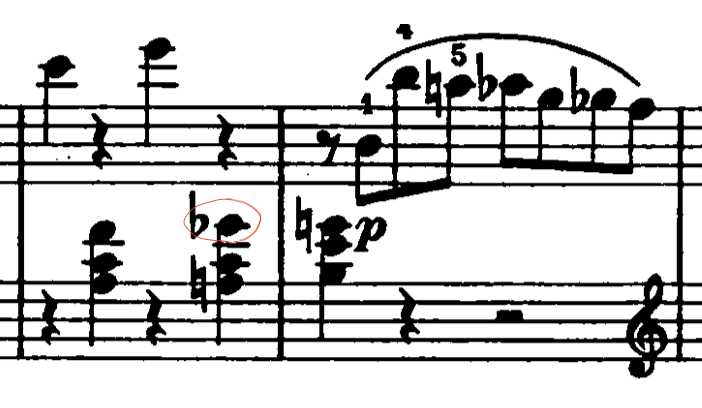}
\caption{Mozart, Sonata no\,14 in c minor, mvt 1, measures 56-57.}
\label{fig:Mozart1}
\end{subfigure}
\begin{subfigure}{0.45\textwidth}
\centering
\includegraphics[height=0.1\textheight, width=0.8\textwidth]{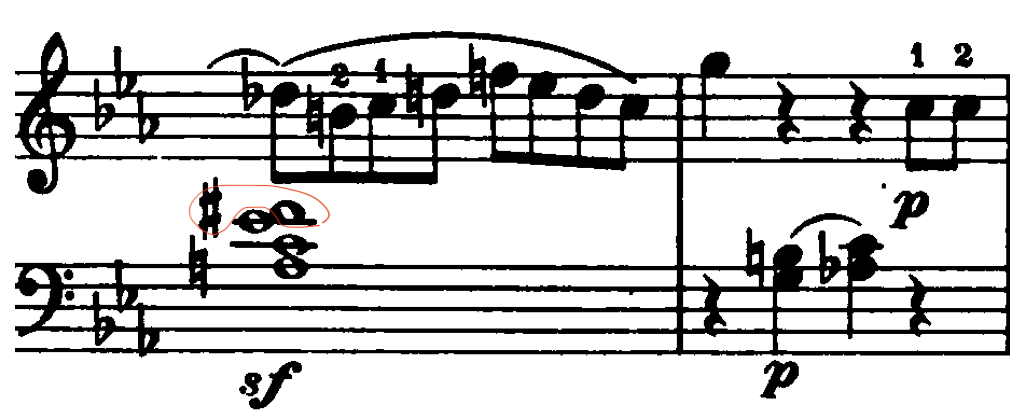}
\caption{Mozart, Sonata no\,14 in c minor, mvt 1, measures 125-126.}
\label{fig:Mozart2}
\end{subfigure}
\caption{Tonal cues in chord spellings in Mozart's Sonata no\,14.}
\label{fig:Mozart}
\end{figure}

In the tonal system, 
fixing a (global) key for a piece defines some default, privileged, names and accidentals.
This rule serves two important purposes in practice for the reader of a music score.
On the one hand, since the default accidentals are not printed, 
the number of symbols displayed on the score is reduced 
and hence the readability is eased.
On the other hand, the key signature immediately poses a tonal context for the piece, and
the presence of other (non default) accidentals constitutes a cue 
of the tonal function of the notes,
which often gives an indication of the composer's intention, 
in particular regarding local key changes. 
For instance, let us consider the last chord of the 56th measure of 
Mozart's $\mathsf{c}$ minor Sonata's 1st movement
(Figure~\ref{fig:Mozart1}), 
which is resolved in the next bar on the second inversion of an $\Ef$ 
Major triad, 
comprising a~$\Gp$ natural 
in the first voice of the left hand. 
By spelling the highest note of the left hand in the chord of interest as~$\Gf$ 
instead of~$\Fs$, 
Mozart chooses not to comply with the principle of selecting an ascending accidental when a voice goes up, 
and rather to express the harmonic function of dominant of the dominant in~$\Ef$, 
which is the new tonality he is heading to at this moment. 
In the same movement, measure 125 (Figure~\ref{fig:Mozart2}), 
a chord composed of the exact same notes is then spelled differently:
an~$\Fs$ 
has replaced the~$\Gf$, 
as the tonal context is shifting back to the main tone 
and the harmonic function has changed to dominant of the dominant in $\mathsf{c}$ minor this time. 
This constitutes one of the numerous examples
(\eg~\cite{Nagel07chromatic}, Chopin's first Ballade or Tristan's chord)
of one chord being spelled differently 
depending on its harmonic nature or tonal function and regardless of the melodic movements of its voices, 
which shows that pitch spelling is indeed revealing in terms of creative intent 
and not only useful to the readability of a piece.
%

There is therefore a 
strong interdependency 
between the problems of \emph{Pitch-Spelling}, 
and (local and global) \emph{Key Estimation} (KE).
%

A large variety of PS algorithms have been proposed, which are
able to guess the spelling of reference corpora
with a high degree of accuracy.
Several such algorithms have been designed according to musicological criteria, 
\florent{revise / PKspell}
selecting note spellings based on the analysis of voice-leading, 
interval relationships and local keys~\cite{temperley2004cognition,chew2005real,meredith2006ps13},
and a principle of parsimony 
(minimisation of the number of accidentals)~\cite{cambouropoulos2003pitch}. 
Some procedures, also based on musicological intuitions, 
reduce PS to optimisation problems in appropriate data structures, 
\eg the Euler lattice~\cite{honingh2009compactness} 
or weighted oriented graphs~\cite{wetherfield2020minimum}. 
%
In other approches, datasets of music scores 
are used to train models such as HMM~\cite{teodoru2007pitch}
or RNN~\cite{foscarin2021pkspell} for PS.
The system in the latter reference, estimates, in addition to PS, 
a key signature for the input piece, with a high accuracy.

In this paper, sharing with~\cite{foscarin2021pkspell} 
the goal of joint Pitch Spe\-ling 
and Key Estimation from MIDI data, 
we generalise the KE problem
from global key signature 
to global and local keys.
Our approach is algorithmic, following the very simple and intuitive idea
to minimise the number of accidental symbols 
\emph{as they would appear on the engraved score}. 
Similar criteria of parsimony are also found 
\eg in~\cite{cambouropoulos2003pitch}.
A difference is that we are applying this principle in a tonal context: 
assuming a key for the piece to spell, 
we count the accidentals for the possible spellings 
according to the \emph{rules of engraving}, 
\ie considering that the accidentals defined by the key signature are not printed, 
and that a printed accidental holds for the embedding measure.

Our procedure works in two steps.
Assuming that the MIDI input is divided into measures, 
we first estimate, 
for each key in a given set, 
the least number of printed accidentals for all possible spelling in all measures.
We use this information in order to estimate a local key for each measure, 
and each possible global key.
Then, in a second step, the above evaluation of the quality of spellings
is refined by taking into account 
(in addition to the number of printed accidentals) 
the proximity of spellings to the evaluated local keys.
The estimated global key is the one giving the best evaluation 
(cumulated for all measures), and the spelling chosen is the one computed 
for this global key.

The assumption of a prior division into measures 
is not required in the above cited papers, 
either those following algorithmic or ML approaches.
The algorithmic papers rather use a sliding window of parametric size.
With that respect, our procedure applies to more restricted cases.
%
However, this assumption makes sense when dealing with quantized MIDI data, 
in particular in the last step of a music transcription framework, 
after rhythm quantization.
Also, the estimated global and local keys, 
which are somehow a side effect of our PS procedure, 
may be useful, as descriptors, in other tasks of processing of
MIDI data for which metrical information is known.
Finally, apart from the boundaries of measures, 
we do not need other information such as note durations or voice separation.

To summarize our contributions, we propose an original and somewhat naive approach to two old problems, which combines them and obtains very good results on challenging datasets.
In Section~\ref{sec:prelim}, we present the preliminary notions
used to state the problems of PS and KE. 
The reader well versed into these problems may skip this section.
We detail our procedure in Section~\ref{sec:method} and present in Section~\ref{sec:evaluation} its evaluation on two datasets
(one monophonic and one piano), for a total of 216 464 notes,
which has given very good results.


\section{Preliminaries}\label{sec:prelim}
We call \emph{part} a polyphonic sequence of notes, organized in measures (bars).
Typically, it shall represent one staff in CWN.
Every note is given by values of pitch, onset and duration.
The representation of these values is described in the two next subsections, 
before another subsection treating the subject of keys and signatures.

\subsection{Pitch representations}
There are two classical alternative representations 
for the note \emph{pitch}, 
distinguishing the input and output of a pitch-spelling algorithm: 
\begin{itemize}
\item 
a \emph{MIDI} value, in 0..128, 
which is a number of semi-tones~\cite{MIDI}, 
\item 
a \emph{spelling}, made of:
\begin{itemize}
\item a note \emph{name} in $\Ap, \ldots, \Gp$,
\item a symbol of \emph{accidental}, amongst 
\musNatural\xspace (natural), 
\musFlat\xspace (flat), 
\musDoubleFlat\xspace (double flat), 
\musSharp\xspace (sharp), 
\musDoubleSharp\xspace (double sharp), 
\item an octave number in -2\ldots 9.
\end{itemize}
\end{itemize}
The lowest MIDI value 0 corresponds to $\Cn${-1}
($\Bs${-2}), 
and the highest one, 128, is $\Gs 9$.
The 88 keys of a piano correspond
to the MIDI numbers~21 ($\An 0$)
to~96 ($\Cn 7$).
A MIDI value modulo~12, is called \emph{pitch class}.
The pitch class of a note~$\nu$ is denoted by $\pc(\nu)$.

\noindent 
Every note name is associated a unique pitch class: 
0 for $\Cp$ to~11 for $\Bp$.
An accidental symbol acts as a pitch class modifier: 
\musDoubleFlat, 
\musFlat, 
\musNatural, 
\musSharp, 
\musDoubleSharp, 
respectively add -2, -1, 0, 1, and 2 to the pitch class 
of the note name component in a spelling.
In the following,
the symbol \musNatural\xspace is sometimes omitted, 
\ie written as a space.
Altogether, with the octave component, 
this principle permits to associate a unique MIDI value to a given spelling.
In the other direction, there are several alternative valid spellings  
for a given MIDI value, 
as summarized in Figure~\ref{fig:enharmonics} for the 12 pitch classes.

\begin{figure}
\begin{center}
\begin{tabular}{|c|ccc|}
\hline
pitch class & spelling$_1$ & spelling$_2$ & spelling$_3$\\
\hline
  0 & $\Dff$   & $\Cp$ & $\Bs$ \\
  1 & $\Df$    & $\Cs$ & [$\Bss$] \\
  2 & $\Eff$   & $\Dp$ & $\Css$ \\
  3 & $[\Fff]$ & $\Ef$ & $\Ds$ \\
  4 & $\Ff$    & $\Ep$ & $\Dss$ \\
  5 & $\Gff$   & $\Fp$ & $\Es$ \\
  6 & $\Gf$    & $\Fs$ & [$\Ess$] \\
  7 & $\Aff$   & $\Gp$ & $\Fss$ \\
  8 & $\Af$    & $\Gs$ &        \\
  9 & $\Bff$   & $\Ap$ & $\Gss$ \\
 10 & [$\Cff$] & $\Bf$ & $\As$ \\
 11 & $\Bf$    & $\Bp$ & $\Ass$ \\
 \hline
 \end{tabular}
\end{center}
\caption{Enharmonic spellings for each pitch class.}
\label{fig:enharmonics}
\end{figure}
Note that, for a every pitch class, the 2 or 3 alternative spellings 
of Figure~\ref{fig:enharmonics} have different names.
Hence, for the purpose of finding a spelling for 
a pitch of MIDI value $m$, is is sufficient to chose 
one of the 2 or 3 possible names for $m$ modulo 12.
The corresponding accidental symbol and octave number can 
then be deduced from the name chosen for $m$.
\begin{figure}
\includegraphics[width=0.46\textwidth]{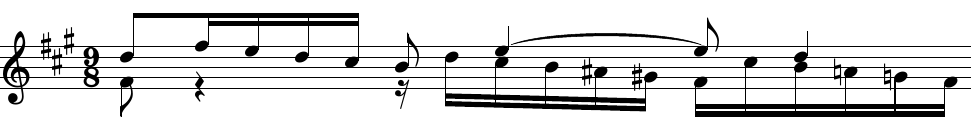}
\caption{Bach, Fugue in A Major BWV864, measure 33, rh.} 
\label{fig:BWV864}
\end{figure}
\begin{example} \label{ex:running}
Figure~\ref{fig:BWV864} presents 
the right-hand part in measure 33 of the Fugue in A Major BWV864
of J.S. Bach.
The MIDI values of the two notes on the first beat of the measure are 66, 
with possible spellings
either $\Fs 4$ 
or $\Gf 4$, 
and 74, 
with possible spellings either
$\Dp 5$, 
or $\Css 5$, 
or $\Eff 5$. 
Here the chosen spelling does not induce any accidental, 
as $\Fs$ is included in the key signature.
An alternative spelling to the one on the score for these two notes could thus be 
$\Gf$, 
$\Dp$ 
but it would generate an added accidental on the $\Gp$ 
since the signature does not contain any $\Gf$; 
hence we observe the importance of the key signature for pitch spelling.
\end{example}
%

\subsection{Time information}
\medskip
In this paper, we assume that the \emph{onset} and \emph{duration} values of notes
are rational numbers, expressed in fraction of a measure. 
The choice of a time unit is not really relevant in this work.
The informations related to time that is important in our procedure are actually:
\begin{itemize}
\item an ordering and equality relation on onsets, 
for sorting the notes in input and detecting notes starting simultaneously, 
\item the number of measure to which a note belongs, 
\ie the detection of bar changes in the flow of input notes. 
\end{itemize}

\noindent We call \emph{grace-note} a note of theoretical duration zero. 
We use this term generically for a note which is part of an ornament
(appoggiatura, gruppetto, mordent, trill \etc).
Two notes are called \emph{simultaneous} if they have the same onset 
and are not grace notes.
This definition can correspond to notes involved in the same chord, 
or notes in different voices and starting simultaneously.
\begin{example}
For instance, the two first notes of Figure~\ref{fig:BWV864} do not really constitute a chord 
but they are \emph{simultaneous} since they share the same onset: 0. 
The time signature for the measure is~9/8, 
hence the two first notes $\Fs$ and $\Dn$ both have a duration of~$\frac{1}{9}$ bar, 
whereas the next semi-quaver $\Fs 5$, 
starting at the onset~$\frac{2}{9}$,
has a duration of~$\frac{1}{18}$.
\end{example}
%

\subsection{Key Signature and Key}
\florent{too basic ?}
A Key Signature (KS) is an integral number $k$ between -7 and 7, 
which indicates that by default, 
$|k|$ note names shall be altered 
by a 
sharp accidental symbol (when $0 < k \leq 7$) 
or a 
flat accidental (when $-7 \leq k < 0$).
The names of the notes altered are defined according to the order of fifths: 
$\Fs, \Cs, \Gs, \Ds, \As, \Es, \Bs$ for the sharps ($k$ positive)
and 
$\Bf, \Ef, \Af, \Df, \Gf, \Cf, \Ff,$ for the flats ($k$ negative).
In a score, a KS indication is placed at the beginning of a part and will
influence the display of the spelling of notes in every measure,
as explained in Section~\ref{sec:convention}.
The KS can be changed during a part.

From the notational point of view (\ie for engraving), 
the choice of KS will change drastically the way how the notes are displayed on the score, 
as it influences the number of accidental symbols printed, 
hence the readability of the score.

From a musical point of view, a KS together with a \emph{mode},  
defines a \emph{global key} $K$ for a piece, 
which is a central notion in tonal music. 
Indeed, the key identifies a diatonic scale, 
whose first note (amongst seven notes), 
called the \emph{tonic} note, 
represents the main tonal focus of the piece.
%
%
\newcommand{\musNat}{\;}
\begin{figure*}
\begin{tabular}{c|ccc|ccccccccc|ccc|}
key signature & $-7$ & $-6$ & $-5$ & $-4$ & $-3$ & $-2$ & $-1$ & 
$0$ & $1$ & $2$ & $3$ & $4$ & $5$ & $6$ & $7$ \\
\hline
major mode & 
$\Cf$ & $\Gf$ & $\Df$ & $\Af$ & $\Ef$ & $\Bf$ & $\Fp$ & 
$\Cp$ & $\Gp$ & $\Dp$ & $\Ap$ & $\Ep$ & $\Bp$ & $\Fs$ & $\Cs$\\
minor modes & 
$\Af$ & $\Ef$ & $\Bf$ & $\Fp$ & $\Cp$ & $\Gp$ & $\Dp$ & 
$\Ap$ & $\Ep$ & $\Bp$ & $\Fs$ & $\Cs$ & $\Gs$ & $\Ds$ & $\As$\\[2pt]
$s_0:$ 
\begin{tabular}{|c|c|}
\hline
$\Cp$\\
\hline
$\Dp$\\ 
\hline
$\Ep$\\ 
\hline
$\Fp$\\ 
\hline
$\Gp$\\ 
\hline
$\Ap$\\ 
\hline
$\Bp$\\ 
\hline
\end{tabular}
$\to$
& 
\begin{tabular}{|c|c|}
\hline
\musFlat\\
\hline
\musFlat\\ 
\hline
\musFlat\\ 
\hline
\musFlat\\ 
\hline
\musFlat\\ 
\hline
\musFlat\\ 
\hline
\musFlat\\ 
\hline
\end{tabular} 
& 
\begin{tabular}{|c|c|}
\hline
\musFlat\\
\hline
\musFlat\\ 
\hline
\musFlat\\ 
\hline
\musNat\\ 
\hline
\musFlat\\ 
\hline
\musFlat\\ 
\hline
\musFlat\\ 
\hline
\end{tabular} 
& 
\begin{tabular}{|c|c|}
\hline
\musNat\\
\hline
\musFlat\\ 
\hline
\musFlat\\ 
\hline
\musNat\\ 
\hline
\musFlat\\ 
\hline
\musFlat\\ 
\hline
\musFlat\\ 
\hline
\end{tabular} 
& 
\begin{tabular}{|c|c|}
\hline
\musNat\\
\hline
\musFlat\\ 
\hline
\musFlat\\ 
\hline
\musNat\\ 
\hline
\musNat\\ 
\hline
\musFlat\\ 
\hline
\musFlat\\ 
\hline
\end{tabular} 
& 
\begin{tabular}{|c|c|}
\hline
\musNat\\
\hline
\musNat\\ 
\hline
\musFlat\\ 
\hline
\musNat\\ 
\hline
\musNat\\ 
\hline
\musFlat\\ 
\hline
\musFlat\\ 
\hline
\end{tabular} 
& 
\begin{tabular}{|c|c|}
\hline
\musNat\\
\hline
\musNat\\ 
\hline
\musFlat\\ 
\hline
\musNat\\ 
\hline
\musNat\\ 
\hline
\musNat\\ 
\hline
\musFlat\\ 
\hline
\end{tabular} 
& 
\begin{tabular}{|c|c|}
\hline
\musNat\\
\hline
\musNat\\ 
\hline
\musNat\\ 
\hline
\musNat\\ 
\hline
\musNat\\ 
\hline
\musNat\\ 
\hline
\musFlat\\ 
\hline
\end{tabular} 
& 
\begin{tabular}{|c|c|}
\hline
\musNat\\
\hline
\musNat\\ 
\hline
\musNat\\ 
\hline
\musNat\\ 
\hline
\musNat\\ 
\hline
\musNat\\ 
\hline
\musNat\\ 
\hline
\end{tabular} 
& 
\begin{tabular}{|c|c|}
\hline
\musNat\\
\hline
\musNat\\ 
\hline
\musNat\\ 
\hline
\musSharp\\ 
\hline
\musNat\\ 
\hline
\musNat\\ 
\hline
\musNat\\ 
\hline
\end{tabular} 
& 
\begin{tabular}{|c|c|}
\hline
\musSharp\\
\hline
\musNat\\ 
\hline
\musNat\\ 
\hline
\musSharp\\ 
\hline
\musNat\\ 
\hline
\musNat\\ 
\hline
\musNat\\ 
\hline
\end{tabular} 
& 
\begin{tabular}{|c|c|}
\hline
\musSharp\\
\hline
\musNat\\ 
\hline
\musNat\\ 
\hline
\musSharp\\ 
\hline
\musSharp\\ 
\hline
\musNat\\ 
\hline
\musNat\\ 
\hline
\end{tabular} 
& 
\begin{tabular}{|c|c|}
\hline
\musSharp\\
\hline
\musSharp\\ 
\hline
\musNat\\ 
\hline
\musSharp\\ 
\hline
\musSharp\\ 
\hline
\musNat\\ 
\hline
\musNat\\ 
\hline
\end{tabular} 
& 
\begin{tabular}{|c|c|}
\hline
\musSharp\\
\hline
\musSharp\\ 
\hline
\musNat\\ 
\hline
\musSharp\\ 
\hline
\musSharp\\ 
\hline
\musSharp\\ 
\hline
\musNat\\ 
\hline
\end{tabular} 
& 
\begin{tabular}{|c|c|}
\hline
\musSharp\\
\hline
\musSharp\\ 
\hline
\musSharp\\ 
\hline
\musSharp\\ 
\hline
\musSharp\\ 
\hline
\musSharp\\ 
\hline
\musNat\\ 
\hline
\end{tabular} 
& 
\begin{tabular}{|c|c|}
\hline
\musSharp\\
\hline
\musSharp\\ 
\hline
\musSharp\\ 
\hline
\musSharp\\ 
\hline
\musSharp\\ 
\hline
\musSharp\\ 
\hline
\musSharp\\ 
\hline
\end{tabular} 
\end{tabular}
\caption{Key signatures and keys.}
\label{fig:tons}\label{fig:keys}
\end{figure*}
In Figure~\ref{fig:keys}, 
we describe 15 key signatures and the tonic of associated keys.
Keys defined by the same KS and different modes are called \emph{relative}.

Additionally to the global key of a piece (or part), 
some alternative \emph{local keys} can be identified for extracts of 
the piece, and might diverge from the global key through \emph{modulations}.
In this work, we shall estimate a {local key} for each measure in a part, 
and this information is used for pitch spelling (Section~\ref{sec:algo-part}).

We shall consider below, in our joint Pitch Spelling and Key Estimation procedure, 
the \emph{major} mode, and three minor modes: 
\emph{harmonic minor}, \emph{melodic minor} 
(also called \emph{ascending minor})
and \emph{natural minor}
(also called \emph{descending minor} or \emph{Aeolian}).
The three minor modes differ by the alteration of 
some degrees in the scale, notably the leading tone (or subtonic in the case of the natural minor scale) and the submediant with accidentals not in the key signatures: 
seventh degree (raised) for \emph{harmonic minor}, 
and sixth and seventh degrees (raised) for \emph{melodic minor}.

Only the {major} and {harmonic/natural minor} are used for global Key Estimation.
The melodic minor mode is only used for local Key Estimation, 
see Section~\ref{sec:algo-part} and Section~\ref{sec:discussion}
for a discussion on occurrences of these modes in pieces used for evaluation.
%
%
We shall consider a measure of distance between keys 
defined by Gottfried Weber in~\cite{Weber}
(see Appendix~\ref{app:Weber}),
see also~\cite{Feisthauer20smc} for other use of 
Weber's table in the context of Key Estimation.

In theory, 
the list of key signatures can be extended on the right and on the left, 
respectively through double sharps and double flats.
For instance, with $k=8$ ($\Gs$ major), 
$\Fp$ is altered with a double sharps ($\Fss$), 
with $k=9$ ($\Ds$ major), 
$\Fp$ and $\Cp$ are altered with double sharps ($\Fss$, $\Cff$), \etc. 
We do not consider the case of extended KS in this work, 
as they are very rarely found.

In both major and minor modes,
the keys associated with key signatures 
respectively 
-7 and 5, 
-5 and 7, and 
-6 and 6 
have tonics with the same pitch class, but different names.
These keys are called \emph{enharmonic}.
They corresponds to the 3 firsts and 3 lasts columns in Figure~\ref{fig:keys}.
Melodies written in either of two enharmonic keys
cannot be distinguished by hear (in equal temperaments),
and moreover, changing a key for its enharmonic preserves the intervals.
Therefore, spelling in one or the other of enharmonic keys 
is essentially a matter of choice.
Usually, the keys with KS~5 (5~sharps, \eg $\Bp$ major) 
or KS~{-5} (5~flats, \eg $\Df$ major) are preferred
over their enharmonic equivalents
with respectively 
KS~{-7} (\eg $\Cf$ major) 
and KS~7 (\eg $\Cs$ major).
But it is not always the case, 
for instance the 
Prelude and Fugue, BWV~848,
of Bach's Well-Tempered Clavier
are in $\Cs$ major.


\section{Algorithms}\label{sec:method}\label{sec:algo}
We present in this section the procedure that we are using 
to estimate, in the same procedure, best spellings, a global key 
and one local key per measure, for a given sequences of input notes.
Roughly, it works measure by measure, 
by comparing the number of accidentals in 
all spellings for all candidate global keys in a given set.
This exhaustive comparison is done through dynamic programming techniques, 
based on the principle used to decide which accidentals must be printed
or not in CWN (presented in next Section~\ref{sec:convention}).
The estimated local keys are used to refine the selection of spellings 
in each measure, in case of ties.

\subsection{Notational convention modulo}\label{sec:convention}
In order to lighten the notations, and ease the readability, 
some accidental symbols are not printed in scores in CWN.
Following a principe of parsimony, 
the notational conventions are roughly that: 
the accidentals in the key signature are omitted by default, 
and the other accidentals need not be repeated in the same measure.
There is moreover a restriction to this rule, summarised as follows
in~\cite{Gould11Notation}:
\emph{It applies only to the pitch at which it is written:
each additional octave requires a further accidental.}

Let us present below 
a relaxed version of this convention, without the above restriction
for octaves.
Somehow, our approach amounts to reasoning modulo 12, 
replacing the notion of pitch by the notion of pitch class.
This application of the principe of parsimony is more suitable 
to the problem of Pitch Spelling~\cite{cambouropoulos2003pitch}, 
which is more concerned in \emph{counting} the accidentals
than in \emph{printing} them.

Formally, the definition of the relaxed convention 
is based on a notion of \emph{spelling state}, or state for short.
Such a state~$s$ is a mapping 
from the note names in $\{ \Ap,\ldots, \Gp \}$
into accidental symbols in 
$\{ \musDoubleFlat, \musFlat, \musNatural, \musSharp, \musDoubleSharp \}$.
\noindent
Let us assume a key signature $k \in \{ -7, \ldots, 7\}$ 
and a measure containing 
a sequence $\nu_1,\ldots, \nu_p$ of $p$ notes, 
enumerated by increasing onsets.
%
An initial state $s_0$ for the measure is built from $k$ 
as expected (see Figure~\ref{fig:tons}):
for $k = 0$,  
$s_0(n) = \musNatural$ for all name $n$, 
for $k = 1$,  
$s_0(\Fp) = \musSharp$ and
$s_0(n) = \musNatural$ for all other name $n$, \etc.
%
For $1 \leq i \leq p$, 
the state $s_i$ is computed by updating the previous
state $s_{i-1}$ as follows, where
$n$ and $a$ are the name and accidental of $\nu_i$:
\begin{enumerate}
\item[$i.$] \label{algo:case-notprinted}
if $s_{i-1}(n) = a$, then $s_{i} = s_{i-1}$, 
and $a$ is \emph{not printed},
\item[$ii.$] \label{algo:case-printed}
otherwise,
$s_{i}(n) = a$,
$s_{i}(n') = s_{i-1}(n')$ for all $n' \neq n$,
and $a$ is \emph{printed}.
\end{enumerate}

\begin{example}
In Figure~\ref{fig:BWV864}, 
for instance, at all onsets before~$\frac{5}{9}$, the spelling state is composed 
of $\Fs$, $\Cs$, $\Gs$ with every other note of the scale being natural. 
On onset $\frac{5}{9}$ however, the state changes for the first time in the measure 
and now contains $\As$ instead of $\An$. 

The next onset also induces a change in the state with $\Gn$ being replaced by $\Gs$, 
which is a note present in the minor ascending melodic mode of $\Bp$ 
(it is interesting to remark that Bach used it in a descending motion, 
 as a pitch spelling process relying too much on motion direction between notes 
 would have failed here). 

The spelling state then stays the same until onset $\frac{15}{18}$ 
with the return of $\An$ 
(belonging to the natural minor mode of $\Bp$) 
and finally $\Gn$ at onset $\frac{16}{18}$, 
hence the measure finishes with the same state it started with.
\end{example}
%

\subsection{Processing one measure in a key}
\label{sec:algo-measure}\label{sec:algo-bar}
Let $K$ be a key with key signature $k \in \{ -7\ldots 7\}$
and let~$\bar{\nu} = \nu_1,\ldots, \nu_p$ be a sequence 
of notes inside a measure, 
with known MIDI values and unknown spellings.
In order to find the spellings for $\bar{\nu}$
with the least number of printed accidentals, 
according to the conventions of Section~\ref{sec:convention}, 
it is fortunately not necessary to enumerate all the $O(3^p)$ 
possible spellings of $\bar{\nu}$.
Indeed, for that purpose, it is sufficient to perform 
a shortest path search, using the state structure of Section~\ref{sec:convention}.
%
Before presenting the details of the search method, 
let us formulate an additional hypothesis, 
that is relevant musically and turned out to be  
important from a combinatoric point of view:
\begin{description}
\item[(\textit{h})]
\emph{two simultaneous notes in the same pitch class must have the same name.}
\end{description}
The hypothesis~(\textit{h}) 
is not a strict notational convention, 
although counter examples are very rare.
However, we do not assume the other direction
(simultaneous notes with same name must be in the same pitch class),
as it can occur fairly frequently in tonal music, at least from the end of the nineteenth century, for instance in 
a dominant ninth chord with an appoggiatura on the ninth
(\eg $\Dp$ $\Fs$ $\Ap$ $\Cp$ $\Fn$ in $\Gp$ major), much appreciated by Ravel, among others.

\begin{example}
\begin{figure}
\includegraphics[width=0.46\textwidth]{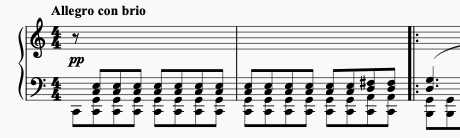}
\caption{Beethoven, Sonata 21 "Waldstein", measures 1-3, lh.} 
\label{fig:Waldstein}
\end{figure}
Without the hypothesis (\textit{h}) stated above, the presence of numerous chords inside of one measure made the complexity of PSE explode in cases such as the beginning of the Waldstein Sonata, displayed in Figure~\ref{fig:Waldstein}. The treatment of simultaneous notes allows the algorithm not to treat all possible spellings for each apparition of doubled notes inside a chord (in this example one of the $\Cp$'s 
in the repeated $\Cp$ major chord need not be treated 
once the other $\Cp$'s spelling has been determined).
\end{example}

In order to ensure hypothesis~(\textit{h}), 
we consider another state variable~$c \in \mathcal{C} = \{ \Ap,\ldots, \Gp \}^{\{ 0,\ldots, 11\}}$
which is a partial mapping of pitch classes 
into note names.
It is used to memorize the name associated to a pitch class
when processing a subsequence of simultaneous notes in $\bar{\nu}$.
The spelling of a note with pitch class~$p$ and note name~$n$
is called \emph{compatible} with $c$ iff
$c(p)$, if defined, is equal to $n$.

\florent{suppr. commentaire size $C$}

\noindent
Let 
$\mathcal{S} = 
\{ \musDoubleFlat, \musFlat, \musNatural, \musSharp, \musDoubleSharp \}^{\{ \Ap,\ldots, \Gp \}}$
be the set of possible values for the state variable $s$ 
defined in Section~\ref{sec:convention},
$I = \{ 1 \leq i < p \mid 
           \nu_{i}\, \mbox{and}\, \nu_{i+1}\,
           \mbox{are simultaneous}\}$,
$\bar{I} = \{ 1, \ldots, p \} \setminus I$, 
and let us consider the following set of configurations:
\[ V = \bigl\{ \< s_0, 0> \bigr\} 
  \cup \mathcal{S} \times \bar{I}
  \cup \mathcal{S} \times \mathcal{S} \times \mathcal{C} \times I.  
  \]
%
The configuration $\< s_0, 0>$ is initial, 
where $s_0$ is the initial state associated to $K$ as in 
Section~\ref{sec:convention}.
In configurations of the form 
$\<s, i> \in \mathcal{S} \times \bar{I}$
we have one state $s$
for a note index $i \in \bar{I}$.
These configurations are dedicated to the processing of single notes.
The other configurations, of the form  
$\<s, t, c, i> \in \mathcal{S} \times \mathcal{S} \times \mathcal{C} \times {I}$
contain additional information in $t$ and $c$
for processing a sub-sequence of simultaneous notes, 
ensuring in particular hypothesis~(\textit{h}).

\noindent
We consider a set of transitions
$E \subset V \times \mathbb{N} \times V$, 
containing the \emph{weighted edges} of one of the following forms:
\begin{eqnarray}
\< s, i-1>        & \xrightarrow[n, a]{w} &\< s', i> 
  \label{tr:single-single}\\
\< s, i-1>        & \xrightarrow[n, a]{w} &\< s', s, c', i> 
  \label{tr:single-simult}\\
\< s, t, c, i-1>  & \xrightarrow[n, a]{w} &\< s', t, c', i> 
  \label{tr:simult-simult}\\
\< s, t, c, i-1>  & \xrightarrow[n, a]{w} &\< s', i> 
  \label{tr:simult-single}
\end{eqnarray}
Every above case means that there exists a spelling of $\nu_i$ 
with name~$n$ and accidental $a$,
which is moreover compatible with $c$ 
in cases~\eqref{tr:simult-simult} and~\eqref{tr:simult-single}.
%
\noindent
In transitions~\eqref{tr:single-single}-\eqref{tr:simult-single},
$s'$ is the update of $s$ as defined 
in Section~\ref{sec:convention} (cases $i$, $ii$).
However, since the purpose of state $s$ 
is no more score engraving like in Section~\ref{sec:convention}, 
but the the search of a pitch spelling, 
we shall distinguish below, 
for the definition of the weight value $w$, 
the cases when the accidental $a$ is \emph{counted} 
from the cases when it is \emph{not counted}
(and not anymore \emph{printed} or \emph{not printed} like in Section~\ref{sec:convention}).

\smallskip
\noindent
The transition~\eqref{tr:single-single} processes the single 
note $\nu_i$ which is not simultaneous with $\nu_{i+1}$.
The condition for counting $a$ 
is the same as the condition for \emph{printing}
in Section~\ref{sec:convention} (cases $i$, $ii$):
\begin{itemize}
\item if $s_{i-1}(n) = a$, then $a$ is \emph{not counted},
\item otherwise, $a$ is \emph{counted}.
\end{itemize}

\smallskip
\noindent
The transition \eqref{tr:single-simult} 
initiates the processing of a subsequence of two or more simultaneous notes, 
when $\nu_i$ is simultaneous with $\nu_{i+1}$.
It makes a copy 
of the current state~$s$ in the second component, 
and create $c' = \{ \< p, n> \}$, with $p = \pc(\nu_i)$. 
The accidental~$a$ is \emph{counted} or \emph{not} 
under the same conditions as for~\eqref{tr:single-single}. 

\smallskip
\noindent
The transition~\eqref{tr:simult-simult}
performs one step of the processing of a subsequence of simultaneous notes, 
again when $\nu_i$ is simultaneous with $\nu_{i+1}$.
It propagates the frozen copy of state $t$ (without modifying it)
and updates $c$ into $c'$ as follows: 
\begin{itemize}
\item if $c(p)$ is defined, then $c' = c$, and $a$ is \emph{not counted},
\item otherwise, $c' = c \cup \{ \< p, n> \}$, 
      and $a$ is \emph{counted} if $t(n) \neq a$.
\end{itemize}

\smallskip
\noindent
The transition \eqref{tr:simult-single}
terminates the processing of a subsequence of simultaneous notes, 
when $\nu_i$ is not simultaneous with $\nu_{i+1}$.
The condition for counting $a$ is the same 
as for~\eqref{tr:simult-simult}:
\begin{itemize}
\item if $c(p)$ is defined, then $a$ is \emph{not counted},
\item otherwise, $a$ is \emph{counted} if $t(n) \neq a$.
\end{itemize}

\noindent
Finally, the weight value of each transition 
in \eqref{tr:single-single}-\eqref{tr:simult-single}
is $w = 0$ if $a$ is \emph{not counted}, 
or if $K$ is in \emph{harmonic minor} or \emph{melodic minor} mode, 
and  $\nu_i$ corresponds to a lead degree in the corresponding scale
and $a$ is the accidental for this degree in the scale.
Otherwise, 
$w = 1$ if $a \in \{ \musFlat, \musNatural, \musSharp \}$,
and $w = 2$ if $a \in \{ \musDoubleFlat, \musDoubleSharp \}$.
The purpose of the above exception for minor modes is to help 
the estimation of keys with minor modes in the next section.

The oriented graph $\mathcal{G} = \<V, E>$ is acyclic.
The vertex $\< s_0, 0>$ is called the \emph{source} of $\mathcal{G}$
(it has no incoming edge) and 
the vertices of $\mathcal{S} \times \{ p \}$ 
are called \emph{targets} of $\mathcal{G}$
(they have no outgoing edge).

Intuitively, a path starting from the source vertex 
$\< s_0, 0>$
and ending with a target vertex of $V$, 
and following the edges of $E$, 
describes a spelling of the sequence of notes~$\bar{\nu}$ in input.
The cumulated sum of the weight values~$w$ of edges involved in such a path
is a count of the accidental symbols printed.

\noindent
The problem of searching for spellings of the notes of~$\bar{\nu}$
with a minimal number of printed accidentals
reduces to the search in~$\mathcal{G}$
of a path with a minimal cumulated weight, 
from the source vertex into a target vertex of~$V$.
That can be done in time $O(|V| + |E|)$
with a greedy Viterbi algorithm~\cite{Huang08advanceddynamic}, 
tagging the vertices of $V$ with cumulated weight values.
It proceeds in a forward way, starting from the source vertex, 
such that the tag of a vertex $v$ is the weight of a minimal path leading to~$v$.

For efficiency, the graph $\mathcal{G}$ is built on-the-fly, 
starting from the source vertex $\< s_0, 0>$, 
and adding vertices along with the computation of their tagging,
following edges complying with the above conditions.
This strategy ensures a pruning of unnecessary search branches, 
making the approach efficient for a use on practical cases, 
as described in Section~\ref{sec:evaluation}.

\subsection{Processing one part}\label{sec:algo-part}
We assume given a sequence of $m$ measures containing notes 
with known MIDI values and unknown spellings.
Let us consider a fixed array of keys $\bar{K} = K_1,\ldots, K_n$.
Our Pitch Spelling approach works by filling a table of dimensions $n \times m$ 
with best spellings for every key in~$\bar{K}$ and every measure, 
using variants of the algorithm of Section~\ref{sec:algo-bar} for each cell.
Then, one estimated global key is selected in~$\bar{K}$,
according to the table content, and the spellings in the corresponding row
of the table can be applied to the input.

\noindent
We proceed in several steps, building actually several tables, 
with several variants for the domain of weight values. 

\subsubsection{step 1: Computation of a first spelling table.}
In the first step, we compute a table $T$ of dimension $n \times m$,
such that the cell $T[i, j]$ contains the cumulated weight (in~$\mathbb{N}$)
of a best spelling, according to the algorithm of Section~\ref{sec:algo-bar}.

\subsubsection{step 2: Estimation of candidate global keys.}
We compute the sum of each row in $T$, 
and save in a list $L$ of candidate global keys
the keys of $\bar{K}$ in \emph{major} and \emph{harmonic or natural minor} modes
with a smallest sum  (there may be ties).

\subsubsection{step 3: Computation of a grid of local keys.}
It is a second table $G$ of dimensions $n \times m$, 
we shall store estimated local keys.
The cell $G[i, j]$ shall contain the local key estimation for 
the measure number~$j$, assuming that $K_i$ is the global key.
The estimation is done column by column
(\ie measure by measure).

For each measure number $0 \leq j \leq m$, we compute 
a ranking $R_j^\mathsf{w}$ of the keys in $\bar{K}$,
according to the values (in~$\mathbb{N}$) of
$T[0, j],\ldots, T[n, j]$, 
the smallest weight value being the best.

Then, for the same $j$ and for each $0 \leq i \leq n$ 
we compute two other rankings $R_{j, i}^\mathsf{p}$ 
and $R_{j, i}^\mathsf{g}$ of $\bar{K}$, 
according to respectively the Weber distance to the previous estimated
local key and the assumed global key
(the smallest distance is the best).
More precisely, the value associated to the key $K_{i'} \in \bar{K}$, 
with $0 \leq i' \leq n$, 
for computing its rank in $R_{j, i}^\mathsf{p}$ is
(see Appendix~\ref{app:Weber} for the Weber table):
\begin{itemize}
\item the Weber distance between $K_{i'}$ and $K_i$ if $j = 0$, 
\item the Weber distance between $K_{i'}$ and 
	  the previous local key in the same row $G[i, j-1]$,
      if $j > 0$.
\end{itemize}
And the value associated to the key~$K_{i'}$
when computing its rank in~$R_{j, i}^\mathsf{g}$ is 
the Weber distance between~$K_{i'}$ and~$K_i$.

\noindent
Finally, for $j$ and $i$, we aggregate the three rankings 
$R_{j}^\mathsf{w}$, 
$R_{j, i}^\mathsf{p}$, 
and $R_{j, i}^\mathsf{g}$
into a unique ranking~$R_{j, i}$,
by comparing, for each $0 \leq i' \leq n$, 
the mean of its three ranks in each ranking --
see~\cite{Demvsar06classifiers} about this method. 
The estimated local key in $G[i, j]$ is the one with the best rank in 
$R_{j, i}$.

In practice, the computation of $G$ can be restricted to the 
rows present in the list $L$ extracted at \emph{step 2}.

\subsubsection{step 4: Computation of a second spelling table.}
In this final step, 
we compute a last table table $U$ of dimension $n \times m$, 
using the same algorithm as in \emph{step 1}, 
but with more involved weight values.

These weight values, 
used to replace the $w \in \mathbb{N}$ in Section~\ref{sec:algo-bar},
are tuplets of integers, with the following components:
\begin{enumerate}
\item \label{weight:accid}
	  the value $w \in \mathbb{N}$ of Section~\ref{sec:algo-bar}, 
\item \label{weight:dist}
      the number of accidentals which do not occur in the 
      scale associated to the estimated local key, 
      for the current processed measure and the current assumed 
      global key (variable $K$ in Section~\ref{sec:algo-bar}),
\item \label{weight:chroma}
      the number of spellings not in the chromatic harmonic scale~\cite{Nagel07chromatic}
      of the estimated local key,
\item \label{weight:color}
	  the number of accidentals with a color different 
	  from the current assumed global key signature $k$
	  (\ie~\musDoubleFlat\xspace or~\musFlat\xspace when $k> 0$
	  and~\musSharp\xspace or~\musDoubleSharp\xspace when $k < 0$),	  
\item \label{weight:cb}
      the number of $\Cf$, or $\Bs$, or $\Ff$, or $\Es$.
\end{enumerate}
The components~\eqref{weight:dist} and~\eqref{weight:chroma}
will second the former weight value~\eqref{weight:accid}
in order to refine the search of best spelling thanks to the 
information gained with the local key estimation in \emph{step 3}.
The two last components \eqref{weight:color} and \eqref{weight:cb}
have been added for the purpose of tie-breacking.

\noindent
We consider two orderings on the domain of above weight values:
\begin{description}
\item $W_{\ell\mathit{ex}}$: the lexicographic comparison 
      of the tuplets with the 5 components \eqref{weight:accid}-\eqref{weight:cb}, 
\item $W_{+}$: the lexicographic comparison of the 4-uplets with, 
	  for first component the sum of \eqref{weight:accid} and \eqref{weight:dist}, 
	  and for next components 
	  \eqref{weight:chroma}, \eqref{weight:color}, and \eqref{weight:cb}
	  respectively.          
\end{description}

Using the same technique as in \emph{step 2}, 
we extract from $L$ one unique estimated global key $K_i \in \bar{K}$, 
using the above refined weight values, 
and apply the spellings found in the row $i$ of the table $U$.

\begin{example}\label{ex:Dijkstra}
A dump of the first table (\emph{step 1}) computed 
when treating the Fugue BWV 864 (see the extract in Figure~\ref{fig:BWV864})
can be found in Appendix~\ref{app:tables}.

Here, 2~global candidates are obtained: 
$\Ap$ major (3 sharps) and $\Fs$ minor (3~sharps) 
as their cumulated row costs are close and inferior to all the others.
Only their corresponding rows will now be computed in the second table 
(\emph{step 3}), 
this time taking into account local tonal analysis to refine the choice between candidate best paths.
The second table is also displayed in Appendix~\ref{app:tables}.

Only 1 global candidate remains at the end: $\Fs$ minor (3~sharps).
Even if the piece is in $\Ap$ major instead of $\Fs$ minor, 
the process found the correct global key signature, 
which is what impacts the pitch spelling. 
This information, along with the local tonalities computed in each measure, 
and used in the second table to choose between paths with the same number of accidents,
was critical in finally selecting a spelling. 
Indeed, for this piece, the algorithm reached an accuracy 
of 100\% compared to the groundtruth\footnote{The score obtained 
for Bach's Fugue BWV 864 (Example~\ref{ex:Dijkstra}) can be found in
\url{https://anonymous.4open.science/r/PSEval-2E27/Results_ASAP/Bach/PSE/BWV864_Fugue.musicxml}}.
\end{example}

\subsection{Rewriting passing notes}
After choosing a spelling with the algorithm of Section~\ref{sec:algo-part}, 
we apply local corrections by rewriting the passing notes, 
using a slight generalisation of the rewrite rules proposed by D. Meredith 
in the original PS13 Pitch-Spelling algorithm~\cite{meredith2006ps13}, 
step 2. 

Every rule applies to a trigram of notes $\nu_0, \nu_1, \nu_2$, 
and rewrites the middle note $\nu_1$, by changing its name.
\newcommand{\rewrite}[5]{{#1} \overset{#2}{\mbox{---}} {#3} 
                              \overset{#4}{\mbox{---}} {#5} }
In Figure~\ref{fig:rewrite}, 
we present the rules for particular cases of notes. 
In general however, the rules are defined by patterns 
comparing the respective note names 
of $\nu_0$, $\nu_1$, and $\nu_2$ (without the accidentals), 
and the difference between their pitch (in number of semitones).
For instance, in the left-hand-side ${\Cp\, \Cf\, \Cp}$ 
of the first rule \emph{broderie~down}, 
$\nu_0$, $\nu_1$, and $\nu_2$ all have the same note name~$\Cp$, 
the difference, in semitons, between $\nu_0$ and $\nu_1$ is $-1$ 
and the difference between $\nu_1$ and $\nu_2$ is $+1$. 
This rule rewrites the middle $\Cf$ ($\nu_1$) 
into~$\Bp$.


\begin{figure}
\[
\begin{array}{lrcl}
\mbox{broderie~down} & 
  \Cp\; \Cf\; \Cp & \to & \Cp\; \Bp\; \Cp\\
\mbox{broderie up} & 
  \Cp\; \Cs\; \Cp & \to & \Cp\; \Df\; \Cp\\
\mbox{descending}_{11} &
  \Cp\; \Cf\; \Ap & \to & \Cp\; \Bp\; \Ap\\
\mbox{descending}_{12} &
  \Cp\; \Cff\; \Af & \to & \Cp\; \Bf\; \Af\\
\mbox{descending}_{21} &
  \Cp\; \As\; \Ap & \to & \Cp\; \Bf\; \Ap\\
\mbox{descending}_{22} &
  \Cp\; \As\; \Af & \to & \Cp\; \Bf\; \Af\\
\mbox{ascending}_{11} &
  \Ap\; \As\; \Cp & \to & \Ap\; \Bf\; \Cp\\
\mbox{ascending}_{12} &
  \Af\; \As\; \Cp & \to & \Af\; \Bf\; \Cp\\
\mbox{ascending}_{21} &
  \Ap\; \Cf\; \Cp & \to & \Ap\; \Bp\; \Cp\\
\mbox{ascending}_{22} &
  \Ap\; \Cf\; \Cs & \to & \Ap\; \Bp\; \Cs\\
\end{array}
\]
\caption{Rewrite rules for passing notes (particular cases).}
\label{fig:rewrite}
\end{figure}

%
The rewrite rules are applied from left to right 
to the sequence spelled notes. 
Note that at each rewrite step, at most one rule can be applied.

\subsection{Deterministic variant} \label{sec:PS13b}\label{sec:PS14}
We propose a variant of the algorithm presented 
in Sections~\ref{sec:algo-bar} and~\ref{sec:algo-part}, 
which is more efficient but less exhaustive.
This variant, called~{PS13b}
(as in "{PS13} with bar info"), 
is very similar to~{PS13}~\cite{meredith2006ps13}, 
except that it uses the information on measures, 
which is assumed available in this paper but not in~\cite{meredith2006ps13}, 
in order to estimate global and local keys.
In~\cite{meredith2006ps13}, that estimation
is done (implicitely) by counting the number of occurrences  
of the (assumed) tonic note in a window whose
optimal size was evaluated manually.

In the algorithm {PS13b}, 
the choice of the spelling 
with name~$n$ and accidental $a$ for the input note $\nu_i$ 
(Section~\ref{sec:algo-bar}, 
 transition rules~\eqref{tr:single-single}-\eqref{tr:simult-single}),
is forced to the (unique) spelling 
in the chromatic harmonic scale of the current key $K$~\cite{Nagel07chromatic}.
Hence, the transitions are deterministic and there is 
no need to search for best spelling in a measure because there is only one.
The rest of the algorithm works as described in Section~\ref{sec:algo-part}.
The complexity of the table construction in this case is
$O(n \times p)$ where $p$ is the total number of notes in input and
$n$ is the number of keys considered.
This complexity is significantly better than the one of the exhaustive algorithm in
Sections~\ref{sec:algo-bar} and~\ref{sec:algo-part}.
In counterpart, some potentially correct spellings will be missed.


\section{Evaluation}\label{sec:evaluation}
\subsection{Implementation}\label{sec:implem}
The algorithms PSE (of Section~\ref{sec:algo-part})
and PS13b (of Section~\ref{sec:PS13b})
have been implemented\footnote{The C++ code as well as the Python evaluation scripts are
publicly accessible at 
\url{https://anonymous.4open.science/r/PSE-DB4D}} 
in \textsf{C++}20.
This language was chosen for the sake of efficiency 
and for integration into larger systems.
The implementation is object oriented, 
with general classes for pitches, keys, \etc,
and data structures specific to the algorithm, 
such as states, bags of best paths and tables.

The input must be provided by a note enumerator, 
associating to each natural number 
a midi pitch, a bar number, and a flag of 
simultaneity (with the next note).
Therefore, our algorithm can be integrated in a larger project. 
This has been done for a MIDI-to-score transcription framework, 
where the timings (in particular the bar boundaries) are computed before pitch spelling.

A Python binding, based on \textsf{pybind11}~\cite{pybind11}, 
was also written and used for evaluation. 
It offers calls (in Python) to the methods of the \textsf{C++} implementation, 
for the various procedures and steps presented in Section~\ref{sec:method}.

For the evaluation, we used the Music21 toolkit~\cite{Cuthbert10music21}, 
in association with the above Python binding.
Music21 parses the MusicXML files in the evaluation datasets 
(see Section~\ref{sec:datasets})
and extracts for each note the information needed by the algorithm:
\begin{itemize}
\item the MIDI key value,
\item the number of the measure the note belongs to,
\item a flag telling whether the note is simultaneous with the next one.
\end{itemize}
This information is fed to the PS procedure, 
and the spellings computed are compared to the ones in the original scores.
Moreover, some scores are produced in output that highlight 
the errors of our procedure with colour codes, 
and mark each measure with the estimated local key\footnote{\label{foot:PSEval}
See 
\url{https://anonymous.4open.science/r/PSEval-2E27} for the evaluation results,
including annotated scores and tabular summaries.}%
\addtocounter{footnote}{-1}\addtocounter{Hfootnote}{-1}.
%
%

%

\subsection{Datasets}\label{sec:datasets}
Two main datasets were used for performance evaluation: 
a monophonic (complex) one, 
originated from the Lamarque-Goudard rhythm textbook~\cite{Lamarque} 
\emph{D'un Rythme à l'Autre}, 
and the ASAP piano dataset~\cite{foscarin2020asap}.

Performances of both algorithms described in Section~\ref{sec:algo},
PSE and PS13b,
were assessed on the integrality of the Lamarque-Goudard dataset, 
containing 250 excerpts, as MusicXML files, from pieces of extremely various styles, 
from Bach and Scarlatti to Wolf, Duparc, Debussy and Ibert.

Evaluation was also executed on 5 separate corpus from the 222 pieces, 
also in MusicXML format, 
of the ASAP piano dataset. 
All Bach preludes and fugues from the Well Tempered Clavier present in ASAP were used, 
except Preludes BWV 856 and 873 for technical reasons. 
All sonata movements by Mozart and Beethoven included in ASAP were also tested, 
as well as the K 475 Fantaisie by Mozart. 
Every one of the 13 Chopin Etudes contained in ASAP, from both opus 10 and 25, 
was used, as well as the 8 Rachmaninov preludes present, from both opus 23 and 32. 
The cumulated total of notes spelled by our two algorithms for this evaluation 
reaches a value of 216 464.

\subsection{Results and discussion}\label{sec:results}\label{sec:discussion}

Experimentations were conducted for several combinations of the weight domains 
in $\{ W_{\ell\mathit{ex}}, W_{+} \}$,
with the best results obtained when using the weights of $W_+$.
The execution time is about 1.68s on average per piece of the evaluation corpus
(subset of Bach WTC), with the exhaustive algorithm PSE presented in Section~\ref{sec:algo-part},
whereas it is only 0.04s on average per piece with the deterministic variant PS13b 
presented in Section~\ref{sec:PS13b}, with results less accurate by more than 1\%.

\begin{table*}[]
\caption{Accuracy results of both pitch spelling algorithms on pieces from widely different styles and correctness of their global key signature estimation, 
          compared to Krumhansl-Schmuckler algorithm's performances (Music 21 implementation).}
\label{table:global}
 \begin{tabular}{||p{3cm}|p{1.5cm}|p{1.5cm}|p{1.5cm}|p{2.2cm}|p{2.2cm}|p{2.2cm}||}
 \hline
   & number of spelled notes & pitch spelling PSE & pitch spelling PS13b & key estimation PSE & key estimation PS13b & key estimation Krumhansl-Schmuckler \\
 \hline
 Bach WTC ASAP corpus &55530& 99.50\% & 98.27\% & 99.09\% & 98.29\%&87.27\%  \\ 
 \hline
 5 movements from Mozart Sonatas present in ASAP&10043& 99.11\% & 97.30\% & 100\% & 100\%&80\%  \\
 \hline
 Fantaisie K. 745 plus 5 movements from Mozart Sonatas & 13830& 97.65\% & 95.97\% & 80\% & 80\%&60\%  \\
 \hline
 33 movements from Beethoven Sonatas & 87292& 97.64\% & 95.65\% & 92.32\% & 95.71\% &66.15\%\\
 \hline
 13 Etudes by Chopin &25103&96.71\%&96.03 \% & 96.15\%&96.15 \%&84.62\% \\
 \hline
 4 Rachmaninov Preludes & 7022& 98.76\% & 97.49\% &100\% & 100\% & 100\% \\ 
 \hline
Lamarque-Goudard  &27687& 98.46\% & 98.23\% & 76.90\% & 74.30\% &  50.60\%\\ 
 \hline
\end{tabular}
\end{table*}

A summary of the evaluation results is presented in Table~\ref{table:global}.
The detailed results 
are also accessible online\footnotemark. 
They are organised by corpus, 
each comprising a folder for each of the algorithm PSE and PS13b.
Results for alternative versions of the algorithms 
corresponding to different ways of combining weights to compute costs
(either lexicographically or additively) are also included).
Every folder contains a table sumarizing the results obtained on all the pieces in the considered corpus. 
In addition to the tables, folders relating to the Bach, Beethoven and Lamarque-Goudard corpora 
include the annotated Music\-XML scores of the treated pieces, 
where the spelling errors are anotated with color codes, 
and green notes indicate an initial error corrected by the final rewriting. 
The anotations also include the global key estimation, 
and local key estimations for each measure. 
The score obtained after the execution of PSE on Bach's Fugue BWV 893 in 
$\mathsf{b}$~minor is included in Appendix~\ref{app:score} as an example.

Regarding global and local tonality estimation, 
our algorithm achieves very good results 
when we compare key signatures together 
but tends to prefer minor tones to their major relatives. 
This can be explained by the large number of notes a minor tonality possesses 
in our acceptation, as we accept spelling from both harmonic and natural minor modes, 
as well as the ascendant melodic one. 
The only error of global tone estimation on our whole Well-Tempered Clavier dataset 
is directly due to this tendency: 
the presence of natural $\Bp$'s in the BWV 870 prelude ($\Cp$ major of book 2), 
in a piece where flat $\Bp$s are also numerous due to modulations to $\Fp$ major, 
$\Dp$ minor \etc., did not prevent our algorithm 
from estimating the piece as written in $\Dp$ minor, 
because these natural $\Bp$'s were interpreted as part 
of the ascendant melodic minor mode of $\Dp$, 
instead of indicators of a $\Cp$ major context. 

About enharmonic tones, 
which are absolutely impossible to distinguish when only pitches and durations of notes are given, 
if the algorithm only proposed the correct global tone or its enharmonic counterpart 
among its global candidates at the end of the first pass, 
and if its final estimated global key is enharmonic to the correct one, 
then the piece is renamed in the enharmonic rival global tone and errors are computed on this version. 
This way of treating that issue is in accordance with the definition 
of a well spelled piece by~\cite{honingh2009compactness}, 
also shared with~\cite{meredith2006ps13}, 
and~\cite{temperley2004cognition}, 
relying on correctness of the intervals of the piece.

\subsection{Comparison with other systems}
On the task of global tonality guessing (KE), we have compared our algorithms' performances to the ones obtained with the Krumhansl-Schmuckler (KS) model for key determination, as implemented 
in the Music21 Python li\-bra\-ry~\cite{Cuthbert10music21}.
This infamous key-finding algorithm computes for every major and minor tonality a correlation coefficient between profile values of the tested key and total durations of their corresponding pitch class in the musical piece considered, it then chooses the best tonality according to the calculated correlation coefficients. It is interesting to note that our algorithms only need to know measure delimitations and not note durations whereas KS uses note durations and does not care about measures. On the whole corpus (both ASAP and Lamarque-Goudard) we attain a 93\% correctness of key signature determination on average, while KS obtains a 75\% accuracy in total. 
\florent{some words on \cite{Feisthauer20smc}?}

The results for global tonality guessing (KE) on the 
La\-mar\-que-Goudard (LG) dataset 
are is rather low, in compared to ASAP.
The LG corpus, extracted from a rhythm textbook, 
consists of 250 short excerpts of longer pieces. 
The pitches and duration of notes appearing in extracts are not necessarily representative of the whole pieces, 
which explains why KS performs rather poorly on it, 
since its correlation coefficients with tonal profiles become erroneous. 
Our algorithms, mainly relying on accidentals number minimization to infer the tonality, therefore prove significantly more robust when it comes to shorter extracts.

We do not provide a comparison table 
on Pitch Spelling results for several reasons.
First, we assume given measure information, 
unlike the algorithms PS13 \cite{meredith2006ps13}, 
CIV \cite{teodoru2007pitch}
or PKSpell \cite{foscarin2021pkspell};
to this respect, a comparison would not be fair.
Second, most of the former evaluations of pitch spelling algorithms 
used as a benchmark the Muse\-data da\-ta\-set 
proposed by D. Meredith~\cite{meredith2006ps13}.
We could not evaluate our algorithms on this dataset
because it does not include the measure boundary information 
required by our procedures.
Since note durations are included in Musedata, 
it would be possible to conduct an evaluation on this dataset
by manually providing a time signature for each of its 216 pieces.
Nevertheless, with success rates (for PS) 
of 99.41\% with PS13~\cite{meredith2006ps13}, 
99.82\% with CIV~\cite{teodoru2007pitch}
and 99.87\% with PKSpell~\cite{foscarin2021pkspell}, 
the remaining room for improvements on this benchmark 
is rather marginal, 
and we preferred to focus on larger datasets like ASAP to evaluate
and improve our algorithms.

The recent system PKSpell~\cite{foscarin2021pkspell},
for which it is reported a 0.13\% error rate on MuseData
(the best results so far),
has also been evaluated on 33 pieces of the challenging 
piano dataset ASAP~\cite{foscarin2020asap}.
It shows on this dataset an accuracy of 96.50\% for the pitch spelling task 
and 90.30\% for key signature estimation.
Since the identities of the 33 pieces for evaluation
are not disclosed in~\cite{foscarin2021pkspell}, 
it is not feasible for this paper to report performances on the exact same pieces. 
However, with an accuracy of 98.19\% on average 
for pitch spelling on 110 pieces (by 5 different composers ranging from Bach to Rachmaninov) 
from the same ASAP dataset, as reported on Table~\ref{table:global},
and 95.58\% for global key signature estimation, 
it is likely that the proposed PSE algorithm (and PS13b) 
should at least have similar performances to PKSpell, if not better.



\section{Conclusion}
We have presented two algorithms, PSE and PS13b 
for joint pitch spelling and estimation of global and local keys,
from MIDI data including information on measure boundaries.
Originally thought to be integrated in a transcription framework, 
these procedures
could also be used in various tasks of music notation processing. 
Since PS13b has proven to be very efficient, it could also be used 
for displaying music notation from MIDI data in real-time.

The evaluation on challenging datasets has shown robust 
results both for pitch spelling and key signature estimation.
Regarding the estimation of keys, there is currently 
a bias towards minor tonalities which are often preferred to their major relative tones that should be corrected. However it does not impact the chosen key signature.

Several directions can be explored in order to improve the current approach, 
such as a refinement of the weight domain for taking into account note durations 
and metric weight (strong or weak beats) when computing the best path in a given measure.
Moreover, in order to improve the accuracy of the tonal analysis, some subtler musical criteria could be implemented, such as cadence detection or chord classification as well as a process to detect justified key signature changes.

Another way to extend our approach to other genres, sometimes less tonal, 
would be to integrate new \emph{modes} into the computation of the PS table.
For instance, one may consider the integration of jazz modes (\eg Ionian, Dorian \etc)
in order to 
tackle the problem of pitch spelling for jazz, which has not been studied at lot in the literature.
This could be of interest in particular for the notation of jazz soli, improvisations, and bass lines 
for instance.



\balance
\bibliography{references}

\clearpage
\onecolumn
\appendix
\pagenumbering{alph}

\section{Appendix}
In this appendix, we present some details about the procedure 
(Section~\ref{app:Weber}) and two samples of evaluation results, 
for illustration.
The complete evaluation results may be found at
\url{https://anonymous.4open.science/r/PSEval-2E27}

\subsection{Table of Weber} \label{app:Weber}
The table of relationship of keys defined in~\cite{Weber}, 
and used in Section~\ref{sec:algo-part}, step 3, 
is displayed below. 
Keys in major mode are uppercase, 
keys in minor mode are lowercase.

\def\AP{\textsf{A}}
\def\BP{\textsf{B}}
\def\CP{\textsf{C}}
\def\DP{\textsf{D}}
\def\EP{\textsf{E}}
\def\FP{\textsf{F}}
\def\GP{\textsf{G}}
\def\aP{\textsf{a}}
\def\bP{\textsf{b}}
\def\cP{\textsf{b}}
\def\dP{\textsf{d}}
\def\eP{\textsf{e}}
\def\fP{\textsf{f}}
\def\gP{\textsf{g}}
\def\AS{\textsf{A}\musSharp} 
\def\BS{\textsf{B}\musSharp}
\def\CS{\textsf{C}\musSharp}
\def\DS{\textsf{D}\musSharp}
\def\ES{\textsf{E}\musSharp}
\def\FS{\textsf{F}\musSharp}
\def\GS{\textsf{G}\musSharp}
\def\aS{\textsf{a}\musSharp} 
\def\bS{\textsf{b}\musSharp}
\def\cS{\textsf{c}\musSharp}
\def\dS{\textsf{d}\musSharp}
\def\eS{\textsf{e}\musSharp}
\def\fS{\textsf{f}\musSharp}
\def\gS{\textsf{g}\musSharp}
\def\AF{\textsf{A}\musFlat} 
\def\BF{\textsf{B}\musFlat}
\def\CF{\textsf{C}\musFlat}
\def\DF{\textsf{D}\musFlat}
\def\EF{\textsf{E}\musFlat}
\def\FF{\textsf{F}\musFlat}
\def\GF{\textsf{G}\musFlat}
\def\aF{\textsf{a}\musFlat} 
\def\bF{\textsf{b}\musFlat}
\def\cF{\textsf{c}\musFlat}
\def\dF{\textsf{d}\musFlat}
\def\eF{\textsf{e}\musFlat}
\def\fF{\textsf{f}\musFlat}
\def\gF{\textsf{g}\musFlat}
%
\begin{center}
\scalebox{0.80}{%
\rotatebox{90}{%
\begin{tabular}{|r|l|rrrrrrrrrrrrrrrrrrrrrrrrrrrrrr|}
\hline
KS  &   & -7 & -6 & -5 & -4 & -3 & -2 & -1  &  0 &  1 &  2 &  3 &  4 &  5 &  6 &  7 & -7 & -6 & -5 & -4 & -3 & -2 & -1 &  0 &  1 &  2 &  3 &  4 &  5 &  6 &  7 \\
    &   & \CF & \GF & \DF & \AF & \EF & \BF & \FP & \CP & \GP & \DP & \AP & \EP & \BP & \FS & \CS & \aF & \eF & \bF & \fP & \cP & \gP & \dP & \aP & \eP & \bP & \fS & \cS & \gS & \dS & \aS \\
\hline
-7 & \CF &  0 &  1 &  2 &  2 &  3 &  4 &  4 &  5 &  6 &  6 &  7 &  8 &  8 &  9 & 10 &  1 &  2 &  3 &  3 &  4 &  5 &  5 &  6 &  7 &  7 &  8 &  9 &  9 & 10 & 11 \\
-6 & \GF &  1 &  0 &  1 &  2 &  2 &  3 &  4 &  4 &  5 &  6 &  6 &  7 &  8 &  8 &  9 &  2 &  1 &  2 &  3 &  3 &  4 &  5 &  5 &  6 &  7 &  7 &  8 &  9 &  9 & 10 \\
-5 & \DF &  2 &  1 &  0 &  1 &  2 &  2 &  3 &  4 &  4 &  5 &  6 &  6 &  7 &  8 &  8 &  2 &  2 &  1 &  2 &  3 &  3 &  4 &  5 &  5 &  6 &  7 &  7 &  8 &  9 &  9 \\
-4 & \AF &  2 &  2 &  1 &  0 &  1 &  2 &  2 &  3 &  4 &  4 &  5 &  6 &  6 &  7 &  8 &  1 &  2 &  2 &  1 &  2 &  3 &  3 &  4 &  5 &  5 &  6 &  7 &  7 &  8 &  9 \\
-3 & \EF &  3 &  2 &  2 &  1 &  0 &  1 &  2 &  2 &  3 &  4 &  4 &  5 &  6 &  6 &  7 &  2 &  1 &  2 &  2 &  1 &  2 &  3 &  3 &  4 &  5 &  5 &  6 &  7 &  7 &  8 \\
-2 & \BF &  4 &  3 &  2 &  2 &  1 &  0 &  1 &  2 &  2 &  3 &  4 &  4 &  5 &  6 &  6 &  3 &  2 &  1 &  2 &  2 &  1 &  2 &  3 &  3 &  4 &  5 &  5 &  6 &  7 &  7 \\
-1 & \FP &  4 &  4 &  3 &  2 &  2 &  1 &  0 &  1 &  2 &  2 &  3 &  4 &  4 &  5 &  6 &  3 &  3 &  2 &  1 &  2 &  2 &  1 &  2 &  3 &  3 &  4 &  5 &  5 &  6 &  7 \\
 0 & \CP &  5 &  4 &  4 &  3 &  2 &  2 &  1 &  0 &  1 &  2 &  2 &  3 &  4 &  4 &  5 &  4 &  3 &  3 &  2 &  1 &  2 &  2 &  1 &  2 &  3 &  3 &  4 &  5 &  5 &  6 \\
 1 & \GP &  6 &  5 &  4 &  4 &  3 &  2 &  2 &  1 &  0 &  1 &  2 &  2 &  3 &  4 &  4 &  5 &  4 &  3 &  3 &  2 &  1 &  2 &  2 &  1 &  2 &  3 &  3 &  4 &  5 &  5 \\
 2 & \DP &  6 &  6 &  5 &  4 &  4 &  3 &  2 &  2 &  1 &  0 &  1 &  2 &  2 &  3 &  4 &  5 &  5 &  4 &  3 &  3 &  2 &  1 &  2 &  2 &  1 &  2 &  3 &  3 &  4 &  5 \\
 3 & \AP &  7 &  6 &  6 &  5 &  4 &  4 &  3 &  2 &  2 &  1 &  0 &  1 &  2 &  2 &  3 &  6 &  5 &  5 &  4 &  3 &  3 &  2 &  1 &  2 &  2 &  1 &  2 &  3 &  3 &  4 \\
 4 & \EP &  8 &  7 &  6 &  6 &  5 &  4 &  4 &  3 &  2 &  2 &  1 &  0 &  1 &  2 &  2 &  7 &  6 &  5 &  5 &  4 &  3 &  3 &  2 &  1 &  2 &  2 &  1 &  2 &  3 &  3 \\
 5 & \BP &  8 &  8 &  7 &  6 &  6 &  5 &  4 &  4 &  3 &  2 &  2 &  1 &  0 &  1 &  2 &  7 &  7 &  6 &  5 &  5 &  4 &  3 &  3 &  2 &  1 &  2 &  2 &  1 &  2 &  3 \\
 6 & \FS &  9 &  8 &  8 &  7 &  6 &  6 &  5 &  4 &  4 &  3 &  2 &  2 &  1 &  0 &  1 &  8 &  7 &  7 &  6 &  5 &  5 &  4 &  3 &  3 &  2 &  1 &  2 &  2 &  1 &  2 \\
 7 & \CS & 10 &  9 &  8 &  8 &  7 &  6 &  6 &  5 &  4 &  4 &  3 &  2 &  2 &  1 &  0 &  9 &  8 &  7 &  7 &  6 &  5 &  5 &  4 &  3 &  3 &  2 &  1 &  2 &  2 &  1 \\
-7 & \aF &  1 &  2 &  2 &  1 &  2 &  3 &  3 &  4 &  5 &  5 &  6 &  7 &  7 &  8 &  9 &  0 &  1 &  2 &  2 &  3 &  4 &  4 &  5 &  6 &  6 &  7 &  8 &  8 &  9 & 10 \\
-6 & \eF &  2 &  1 &  2 &  2 &  1 &  2 &  3 &  3 &  4 &  5 &  5 &  6 &  7 &  7 &  8 &  1 &  0 &  1 &  2 &  2 &  3 &  4 &  4 &  5 &  6 &  6 &  7 &  8 &  8 &  9 \\
-5 & \bP &  3 &  2 &  1 &  2 &  2 &  1 &  2 &  3 &  3 &  4 &  5 &  5 &  6 &  7 &  7 &  2 &  1 &  0 &  1 &  2 &  2 &  3 &  4 &  4 &  5 &  6 &  6 &  7 &  8 &  8 \\
-4 & \fP &  3 &  3 &  2 &  1 &  2 &  2 &  1 &  2 &  3 &  3 &  4 &  5 &  5 &  6 &  7 &  2 &  2 &  1 &  0 &  1 &  2 &  2 &  3 &  4 &  4 &  5 &  6 &  6 &  7 &  8 \\
-3 & \cP &  4 &  3 &  3 &  2 &  1 &  2 &  2 &  1 &  2 &  3 &  3 &  4 &  5 &  5 &  6 &  3 &  2 &  2 &  1 &  0 &  1 &  2 &  2 &  3 &  4 &  4 &  5 &  6 &  6 &  7 \\
-2 & \gP &  5 &  4 &  3 &  3 &  2 &  1 &  2 &  2 &  1 &  2 &  3 &  3 &  4 &  5 &  5 &  4 &  3 &  2 &  2 &  1 &  0 &  1 &  2 &  2 &  3 &  4 &  4 &  5 &  6 &  6 \\
-1 & \dP &  5 &  5 &  4 &  3 &  3 &  2 &  1 &  2 &  2 &  1 &  2 &  3 &  3 &  4 &  5 &  4 &  4 &  3 &  2 &  2 &  1 &  0 &  1 &  2 &  2 &  3 &  4 &  4 &  5 &  6 \\
 0 & \aP &  6 &  5 &  5 &  4 &  3 &  3 &  2 &  1 &  2 &  2 &  1 &  2 &  3 &  3 &  4 &  5 &  4 &  4 &  3 &  2 &  2 &  1 &  0 &  1 &  2 &  2 &  3 &  4 &  4 &  5 \\
 1 & \eP &  7 &  6 &  5 &  5 &  4 &  3 &  3 &  2 &  1 &  2 &  2 &  1 &  2 &  3 &  3 &  6 &  5 &  4 &  4 &  3 &  2 &  2 &  1 &  0 &  1 &  2 &  2 &  3 &  4 &  4 \\
 2 & \bP &  7 &  7 &  6 &  5 &  5 &  4 &  3 &  3 &  2 &  1 &  2 &  2 &  1 &  2 &  3 &  6 &  6 &  5 &  4 &  4 &  3 &  2 &  2 &  1 &  0 &  1 &  2 &  2 &  3 &  4 \\
 3 & \fS &  8 &  7 &  7 &  6 &  5 &  5 &  4 &  3 &  3 &  2 &  1 &  2 &  2 &  1 &  2 &  7 &  6 &  6 &  5 &  4 &  4 &  3 &  2 &  2 &  1 &  0 &  1 &  2 &  2 &  3 \\
 4 & \cS &  9 &  8 &  7 &  7 &  6 &  5 &  5 &  4 &  3 &  3 &  2 &  1 &  2 &  2 &  1 &  8 &  7 &  6 &  6 &  5 &  4 &  4 &  3 &  2 &  2 &  1 &  0 &  1 &  2 &  2 \\
 5 & \gS &  9 &  9 &  8 &  7 &  7 &  6 &  5 &  5 &  4 &  3 &  3 &  2 &  1 &  2 &  2 &  8 &  8 &  7 &  6 &  6 &  5 &  4 &  4 &  3 &  2 &  2 &  1 &  0 &  1 &  2 \\
 6 & \dS & 10 &  9 &  9 &  8 &  7 &  7 &  6 &  5 &  5 &  4 &  3 &  3 &  2 &  1 &  2 &  9 &  8 &  8 &  7 &  6 &  6 &  5 &  4 &  4 &  3 &  2 &  2 &  1 &  0 &  1 \\
 7 & \aS & 11 & 10 &  9 &  9 &  8 &  7 &  7 &  6 &  5 &  5 &  4 &  3 &  3 &  2 &  1 & 10 &  9 &  8 &  8 &  7 &  6 &  6 &  5 &  4 &  4 &  3 &  2 &  2 &  1 &  0 \\
\hline
\end{tabular}
} 
} 
\end{center}


\subsection{Execution Tables for the Fugue BWV 864} \label{app:tables}
We display in Figures~\ref{fig:BWV864T0}
and~\ref{fig:BWV864T1} 
the dumps of the two tables computed 
when processing the Fugue BWV 864, 
see Example~\ref{ex:Dijkstra}.


\begin{figure*}
\begin{verbatim}
spelling 813 notes
PSE: first table 
Row Costs:
Gbmajor (6b) cost accid=199 dist=0 chromarm=188 color=10  cflat=25
Dbmajor (5b) cost accid=251 dist=0 chromarm=237 color=19  cflat=46
Abmajor (4b) cost accid=291 dist=0 chromarm=280 color=35  cflat=44
Ebmajor (3b) cost accid=279 dist=0 chromarm=274 color=123 cflat=24
Bbmajor (2b) cost accid=252 dist=0 chromarm=258 color=246 cflat=8
Fmajor  (1b) cost accid=219 dist=0 chromarm=222 color=274 cflat=0 
Cmajor  (0)  cost accid=169 dist=0 chromarm=175 color=344 cflat=0 
Gmajor  (1#) cost accid=123 dist=0 chromarm=130 color=3   cflat=4 
Dmajor  (2#) cost accid=71  dist=0 chromarm=78  color=3   cflat=4
Amajor  (3#) cost accid=38  dist=0 chromarm=43  color=2   cflat=4
Emajor  (4#) cost accid=69  dist=0 chromarm=81  color=1   cflat=4
Bmajor  (5#) cost accid=113 dist=0 chromarm=124 color=1   cflat=4
F#major (6#) cost accid=154 dist=0 chromarm=165 color=0   cflat=0
Ebminor (6b) cost accid=179 dist=0 chromarm=205 color=23  cflat=19
Bbminor (5b) cost accid=220 dist=0 chromarm=257 color=30  cflat=34
Fminor  (4b) cost accid=255 dist=0 chromarm=283 color=76  cflat=29
Cminor  (3b) cost accid=229 dist=0 chromarm=279 color=159 cflat=12
Gminor  (2b) cost accid=204 dist=0 chromarm=262 color=293 cflat=4
Dminor  (1b) cost accid=167 dist=0 chromarm=222 color=293 cflat=0
Aminor  (0)  cost accid=130 dist=0 chromarm=175 color=344 cflat=0
Eminor  (1#) cost accid=110 dist=0 chromarm=130 color=3   cflat=4
Bminor  (2#) cost accid=67  dist=0 chromarm=78  color=3   cflat=4
F#minor (3#) cost accid=33  dist=0 chromarm=43  color=2   cflat=0
C#minor (4#) cost accid=69  dist=0 chromarm=81  color=1   cflat=4
G#minor (5#) cost accid=111 dist=0 chromarm=124 color=1   cflat=5
D#minor (6#) cost accid=145 dist=0 chromarm=165 color=0   cflat=0
\end{verbatim}
\caption{First table computed for processing the Fugue BWV 864 (Example~\ref{ex:Dijkstra}).}
\label{fig:BWV864T0}
\end{figure*}

\begin{figure*}
\begin{verbatim}
Row Costs:
Amajor  (3 sharps) cost accid=38 dist=35 chromarm=3 color=4 cflat=1
F#minor (3 sharps) cost accid=33 dist=21 chromarm=0 color=0 cflat=0
\end{verbatim}
\caption{Second table computed for processing the Fugue BWV 864 (Example~\ref{ex:Dijkstra}).}
\label{fig:BWV864T1}
\end{figure*}

\subsection{Resulting score of the Fugue BWV 893}\label{app:score}
We show below the annotated score of the Fugue BWV 893 from the ASAP dataset, 
obtained in output of PSE, 
with spelling errors automatically annotated in red.
We have chosen another example than the Fugue BWV 864 of the previous section, 
since the latter was spelled with PSE with an accuracy 100\%
(and therefore no error was to be anotated in the output score).

The scores processed with our method for evaluation, with annotations 
(see Table~\ref{table:global} for the list), as well as tables of results,  
can be found at \url{https://anonymous.4open.science/r/PSEval-2E27}

\includegraphics[page=1,scale=0.75]{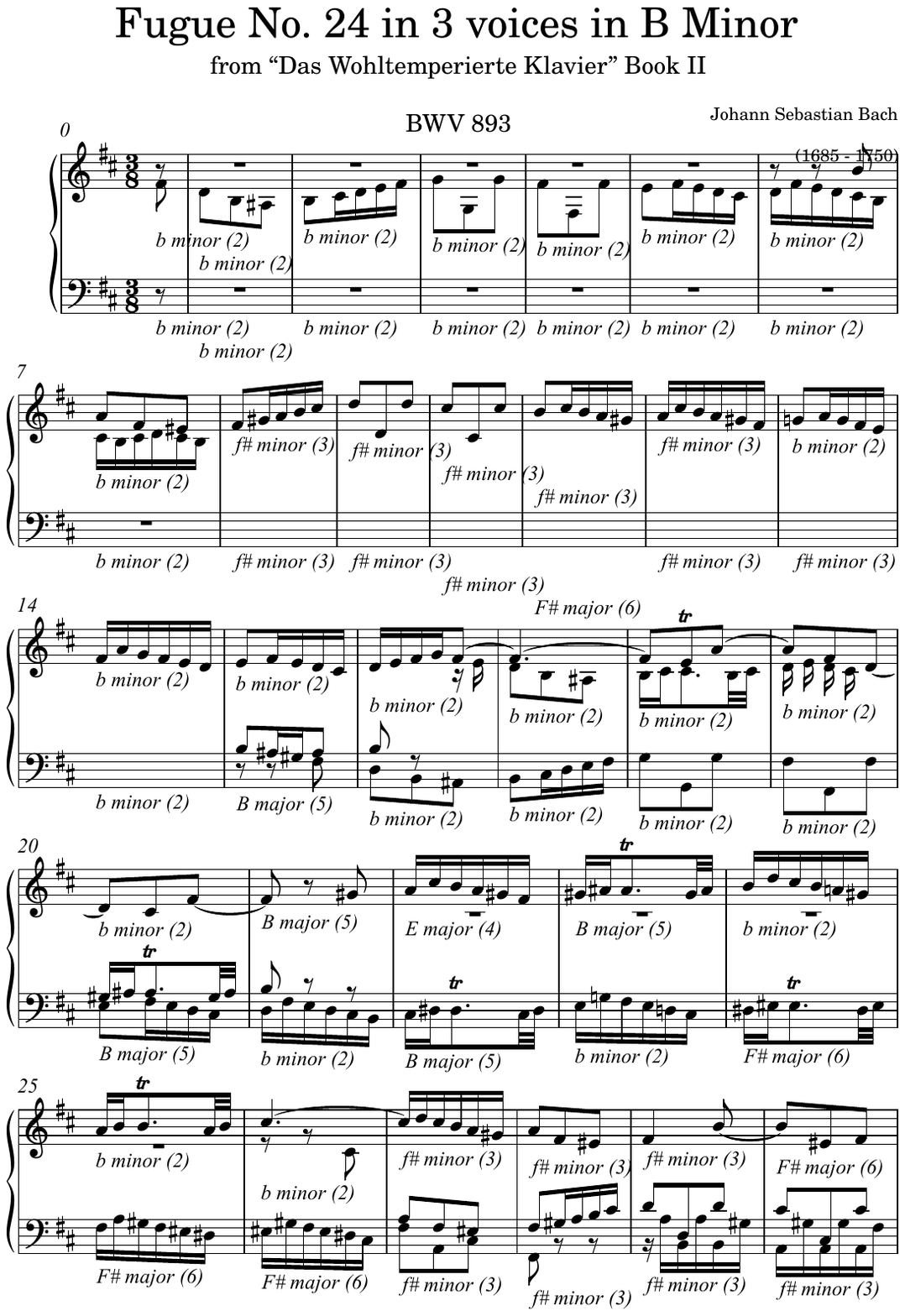}

\includegraphics[page=2,scale=0.8]{BWV893_Fugue.pdf}

\includegraphics[page=3,scale=0.8]{BWV893_Fugue.pdf}

\includegraphics[page=4,scale=0.8]{BWV893_Fugue.pdf}

\end{document}
\endinput